\documentclass[fleqn,usenatbib,letters]{mnras}
\pdfoutput=1                    
\usepackage[T1]{fontenc}
\usepackage[british]{babel}

\usepackage{savesym}
\usepackage{amsmath}
\savesymbol{iint}
\usepackage{txfonts}
\restoresymbol{TXF}{iint}

\usepackage{amssymb,multirow}
\usepackage{bm}
\usepackage{graphicx}
\usepackage{epstopdf}
\usepackage{caption}
\usepackage{float}
\usepackage{hyperref}

\providecommand*{\aap}{{\it A\&A}}
\providecommand*{\mnras}{{\it MNRAS}}

\providecommand*{\grl}{{\it Geophys. Res. Lett.}}

\providecommand*{\pre}   {{\it Physical Review E}}
\providecommand*{\apj}   {{\it Astrophys. J.}}

\renewcommand{\vec}[1]{\bm{#1}}
\renewcommand{\div}[1]{\vec\nabla\bm{\cdot}#1}
\newcommand{\dif}[1]{\vec\nabla^2#1}

\newcommand{\ex}{\vec e_x}
\newcommand{\ey}{\vec e_y}
\newcommand{\ez}{\vec e_z}

\newcommand{\mr}[1]{\mathrm{#1}}

\renewcommand{\thesubsection}{\thesection\alph{subsection}}
\renewcommand{\thesubsubsection}{\thesubsection.\arabic{subsubsection}}
\renewcommand{\thepage}{\thesubsection-\arabic{page}}
\def\rrr#1\\{\par
\medskip\hbox{\vbox{\parindent=2em\hsize=6.12in
\hangindent=4em\hangafter=1#1}}}

\usepackage{color}

\newcommand{\edit}[1]{#1}

\title[Magnetic Turbulence in Rotating, Shearing Flows]{On the Nature of Magnetic Turbulence in Rotating, Shearing Flows}

\author[J. Walker, et al.]{
Justin Walker,$^{1}$\thanks{E-mail: jwwalker2@wisc.edu}
Geoffroy Lesur,$^{2,3}$
and Stanislav Boldyrev$^{1}$
\\
$^{1}$Department of Physics, University of Wisconsin-Madison, 1150 University Avenue, Madison, WI 53706, USA\\
$^{2}$Univ. Grenoble Alpes, IPAG, F-38000 Grenoble, France\\
$^{3}$CNRS, IPAG, F-38000 Grenoble, France
}

\date{\today}

\pubyear{2015}

\begin{document}
\label{firstpage}
\pagerange{\pageref{firstpage}--\pageref{lastpage}}
\maketitle

\begin{abstract} The local properties of turbulence driven by the magnetorotational instability (MRI) in rotating, shearing flows are studied in the framework of a shearing-box model. Based on numerical simulations, we propose that the MRI-driven turbulence comprises two components: the large-scale shear-aligned strong magnetic field and the small-scale fluctuations resembling magnetohydrodynamic (MHD) turbulence. The energy spectrum of the large-scale  component is close to~$k^{-2}$, whereas the spectrum of the small-scale component agrees with the spectrum of strong MHD turbulence~$k^{-3/2}$. While the spectrum of the fluctuations is universal, the outer-scale characteristics of the turbulence are not; they depend on the parameters of the system, such as the net magnetic flux. However, there is remarkable universality among the allowed turbulent states -- their intensity~$v_0$ and their outer scale~$\lambda_0$ satisfy the balance condition $v_0/\lambda_0\sim \mr d\Omega/\mr d\ln r$, where $\mr d\Omega/\mr d\ln r$~is the local orbital shearing rate of the flow.  \edit{Finally, we find no sustained dynamo action in the $\mr{Pm}=1$ zero net-flux case for Reynolds numbers as high as 45\,000, casting doubts on the existence of an MRI dynamo in the $\mr{Pm}\leq 1$ regime. }
\end{abstract}


\section{Introduction}

Magnetorotational instability (MRI), the instability of rotating, shearing flows of plasmas or conducting fluids in the presence of a weak magnetic field \cite[][]{chandra1960,balbus1991}, is thought to play an important role in many natural systems. It was proposed as a mechanism driving angular momentum transport in astrophysical accretion disks \cite[][]{balbus1998}, and it was also studied in the solar dynamo \cite[][]{kagan2014} and geodynamo \cite[][]{petitdemange2008} contexts. The MRI is finally an excellent prototype of subcritical magnetic dynamo action  \cite[][]{rincon2007,herault2011,riols_etal2013}. 

Many analytic, numerical, and laboratory studies have been devoted to the onset of the instability and the resulting magnetic turbulence \cite[e.g.,][]{sisan2004,gissinger2011,seilmayer2014,meheut2015,latter2015}. At present, however, it is hardly possible to address the full-scale dynamics of natural systems exhibiting the MRI, as it incorporates the effects of stratification, global geometry, boundary conditions, etc. However, the local and fundamental properties of MRI driven turbulence may be studied in the framework of reduced models, such as the shearing-box, which isolates  the principal ingredients required for the MRI~\edit{\cite[e.g.,][]{goldreich1965,hawley1995,fromang2007,longaretti2010,lesur2011,riols_etal2015}}. Modelling of the small-scale dynamics produces the angular momentum transport coefficient~$\alpha$ that can be used in global modelling of the disk, star or planet interior~\edit{\cite[e.g.,][]{shakura1973,lesur2010a}}.

Numerical studies of the shearing-box model reveal nontrivial properties of the resulting magnetic turbulence. In the case of zero net magnetic flux through the system -- the so-called MRI-dynamo case -- the turbulence was found to be sustained for magnetic Prandtl numbers exceeding unity, while it was observed to eventually decay for smaller values \citep{fromang2007,balbus2008,riols_etal2013,riols_etal2015}. Larger Reynolds numbers seem to facilitate the MRI dynamo action by lowering the Prandtl number threshold, however, present numerical limitations do not allow one to establish whether this dependence persists at asymptotically large Reynolds numbers. In the case of nonzero net magnetic flux, it was found that the steady state is eventually reached that depends on the value of the flux and also on the magnetic Prandtl number~\cite[][]{longaretti2010}.

In order to understand the numerically observed behaviour it is instructive to understand the properties of the magnetic turbulence that develops in the system. This is the goal of the present letter. It is motivated by several puzzling results obtained in previous works. In particular, previous studies did not find a power-law scaling of the energy spectrum of magnetorotational turbulence \cite[][]{lesur2011}. It remained unknown whether such a system develops a turbulent cascade similar to that found in forced MHD turbulence \cite[e.g.,][]{perez_etal2012,mason2012}, and whether there is any universality among the turbulent states corresponding to different parameter regimes.

In this study we find that in the cases when steady or quasi steady turbulent field is observed, it develops two distinct components. The first component consists of strong magnetic fluctuations almost in the direction of the shear. The spectrum of this component declines as $k^{-2}$; therefore, this component is concentrated at large scales. The remainder of the fluctuations comprise the second, small-scale component that exhibits a turbulent cascade with the shallower spectrum of $-3/2$, similar to that of standard MHD turbulence. The large-scale component of the turbulence plays the role of the guide field for the small-scale component.  We observe that the intensity of the resulting  turbulence depends on the net magnetic flux. However, there is remarkable universality among all the observed turbulent regimes -- the level of turbulence and its outer scale are adjusted in such a way as to ensure that the rate of non-linear interaction is proportional to the shear rate of the background flow.

\section{Numerical setup}

We use the shearing-box or shearing-street approximation of~\cite{goldreich1965}, developed to study the local dynamics of shearing, rotating flow. In this approximation, one considers a small, Cartesian box orbiting at some radius $r_0$ and solves for perturbations from a mean flow. The shearing rate is defined by $q \equiv -( d \ln \Omega / d\ln r)_{r_0}$, where $\Omega(r)$ is the angular speed and the derivative is evaluated at the radial location $r_0$. For a Keplerian flow we have $q=3/2$.
We also denote $\Omega_0= \Omega(r_0)$. The Cartesian coordinates inside the box are chosen in the following way: $x$~is the radial direction, $y$~is the direction of the background velocity, and $z$~is the vertical direction (the direction of $\Omega$).  For simplification, we assume that the flow is incompressible, corresponding to the so-called ``small shearing-box limit;'' the details of the derivation can be found in~\cite{umurhan2004}. When the background mean shear profile is removed, the remaining fluctuating parts of the velocity and magnetic fields obey the following system of equations:
\begin{align}
  D_t \vec{v}  = - (\vec{v} \bm{\cdot} \vec{\nabla})\vec{v} -\nabla P + \vec{B}\bm{\cdot}\nabla{\vec{B}} + \nu \dif{\vec{v}}- \nonumber\\
  -2\vec{\Omega}_0 \bm{\times} \vec{v} + q\Omega_0 v_x \ey  \label{eq:mom},\\
  D_t \vec{B} = \vec{\nabla} \bm{\times} (\vec{v} \bm{\times} \vec{B}) + \eta \dif{\vec{B}} - q\Omega_0 B_x \ey \label{eq:induct},\\
  \div{\vec{v}} = 0 \label{eq:incompress}, \quad \div{\vec{B}} = 0,
\end{align}
\noindent with $D_t\equiv \partial_t-q\Omega_0 x\partial_y$.

\begin{figure}
    \includegraphics{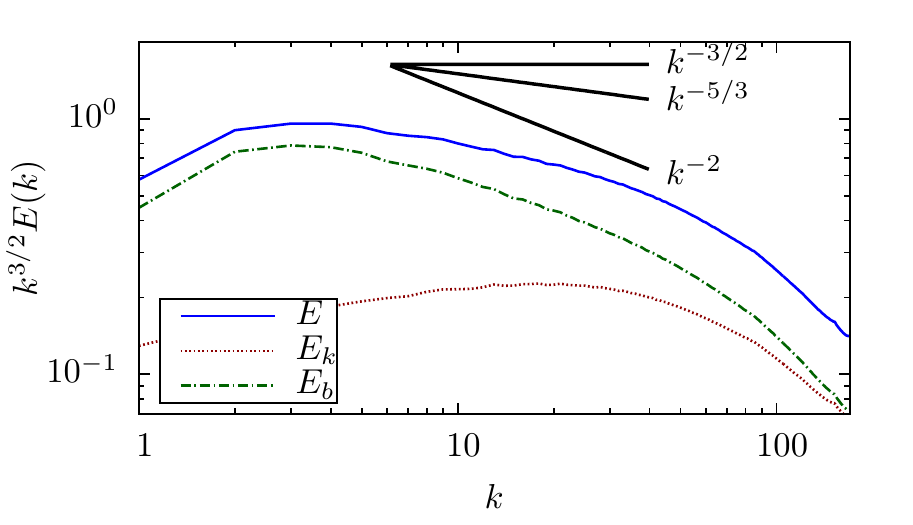}
    \caption{The total energy spectrum, the kinetic energy spectrum, and the magnetic energy spectrum, compensated by $k^{3/2}$, from case~I.}
    \label{fig:spectra}
\end{figure}

\begin{figure}
    \includegraphics{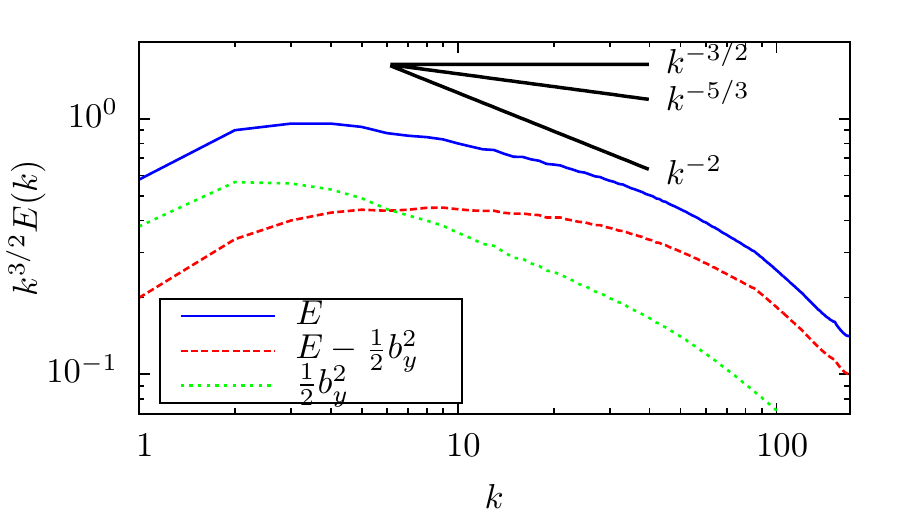}
    \caption{The total energy spectrum, the total energy spectrum without $b_y$, and the spectrum of $b_y$, compensated by $k^{3/2}$, from case~I.}
    \label{fig:spectra2}
\end{figure}

\begin{figure}
  \includegraphics{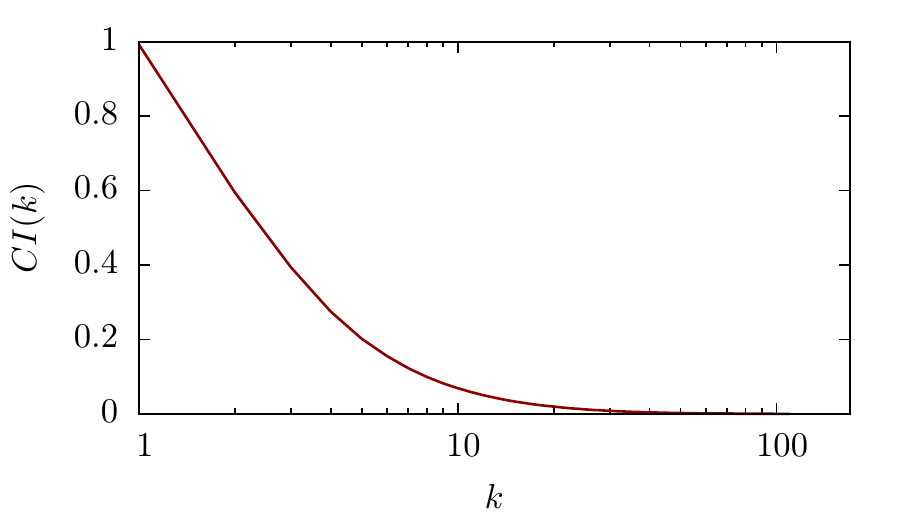}
    \caption{The cumulative energy injection rate at scale~$k$, defined as $CI(k)=\int_k^\infty {\tilde \alpha}(k)dk/\int_1^\infty {\tilde \alpha}(k)dk$, from case~I.}
    \label{fig:injection}
\end{figure}
 
The variables are non-dimensionalized using the inverse of the rotation rate of the disk $t_0 = \Omega_0^{-1}$ 
as the unit of time, the box (disk) height $L_z$ as the unit of length, $\Omega_0 L_z$ as the unit of velocity, and the Alfv\'{e}n speed $v_A = B/\sqrt{4\pi\rho}$ as the unit of magnetic field strength. The density $\rho$ and angular speed $\Omega_0$ are chosen to be unity. In general, when a uniform field $\vec B_0=B_0\ez$ is imposed, we  define the fluctuating part of the magnetic field~${\vec b}$ according to ${\vec B}={\vec B}_0+{\vec b}$.

The relevant dimensionless quantities are the Reynolds number $\mr{Re} = \Omega_0 L_z^2/\nu$, the magnetic Reynolds number $\mr{Rm} = \Omega_0 L_z^2/\eta$, the Elsasser number $\Lambda_{\eta} = B_0^2/\Omega_0\nu$, and the parameter $\beta = \Omega_0^2 L_z^2 /B_0^2$, which measures the strength of the imposed magnetic field and mimics the plasma $\beta$ in vertically stratified disks.  We also introduce the dimensionless transport coefficient $\alpha   \equiv   \langle v_xv_y - b_xb_y \rangle/\Omega_0^2 L_z^2$  and the energy injection rate ${\tilde \alpha} =  q\Omega_0 \langle v_xv_y - b_xb_y \rangle$,
where $\langle \cdot \rangle$ denotes an average performed over the entire volume. From Eqs.~\ref{eq:mom}~and~\ref{eq:induct}, the energy balance equation has the form~\cite[][]{longaretti2010}:
\begin{equation}
\frac{d}{dt}\left< \frac{v^2}{2} +\frac{b^2}{2} \right>=-\nu \big\langle ({\nabla \bm{\times} {\vec v}})^2 \big\rangle -\eta \big\langle ({\nabla \bm{\times} {\vec b}})^2 \big\rangle + {\tilde \alpha}.
\label{energy_balance}
\end{equation}
The Reynolds numbers are chosen to be equal so that the magnetic Prandtl number $\mr{Pm}=\nu/\eta$ is unity.

Equations \eqref{eq:mom}-\eqref{eq:incompress} are solved using the pseudo-spectral code Snoopy~\citep{lesur2007}. Snoopy uses the Fourier transforms of the FFTW 3 library and a low-storage, third-order Runge-Kutta (RK3) scheme for time integration of all terms except the dissipation terms, which are treated implicitly. The shearing street equations solve for fluctuations from a mean background shear, as described above. In order to resolve the solution for a long time, Snoopy periodically remaps the fields every $\Delta t_{\mr{remap}}=|L_y/(q \Omega_0 L_x)|$. Greater detail of the mathematical algorithm is given in~\cite{umurhan2004}. 

In this work, we consider the simulation box with dimensions $L_x:L_y:L_z = 2:4:1$~\citep{bodo2008} and numerical resolution of $N_x\times N_y\times N_z=1024\times 1024\times 512$. All cases have a Reynolds number $\mathrm{Re}=45\,000$. We consider three steady-state cases and one decaying case. 

Case~I has a net flux of $B_0=0.03$, corresponding to~$\beta \approx 1100$. The initial conditions are random fluctuations at large scales that are then evolved over $\sim 50t_0$ to achieve a steady state. case I is this steady state after the  initial growth period, with averages performed over the final~$\sim 50t_0$. The spectra of magnetic and kinetic fluctuations measured in this case are presented in Figs.~\ref{fig:spectra},~\ref{fig:spectra2}, and the energy injection rate ${\tilde \alpha}$ is shown in Fig.~\ref{fig:injection}.

Case~II is another steady-state case that has a weaker net flux of $B_0 = 0.010$ and used a snapshot of case~I as its initial condition. After the steady state had  been reached, the averages were performed over the final~$\sim 20t_0$.

Case~III corresponds to a very weak net flux of $B_0 = 0.005$. To initiate this run, the weak magnetic field was added to the simulations of our zero net flux setup -- case~IV below -- such that $\Lambda_{\eta}\approx 1$. This was so that the linear MRI, which is quenched at high wavenumbers for $\Lambda_{\eta}<1$, would be excited with a minimal injection of energy. We observe that the turbulence reaches a new, lower-energy steady state in this case. Averages were performed over the final~$\sim 100t_0$.

\begin{figure}
  \includegraphics{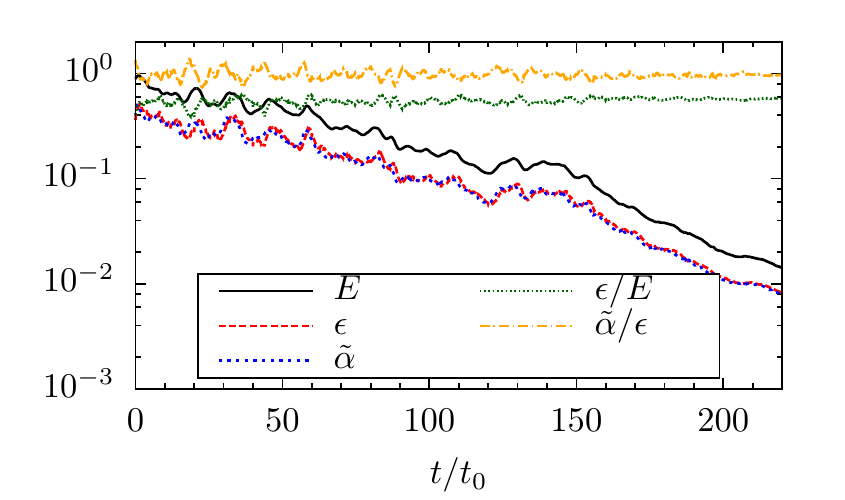}
  \caption{Time history of case~IV showing energy $E$, energy injection rate ${\tilde \alpha}$, and energy dissipation rate $\epsilon$.}
  \label{fig:energy}
\end{figure}

Case~IV is a zero net-flux case that used \edit{as its initial condition a snapshot of case~I in which the mean field was manually zeroed}. We observe that in this case the energy declines very slowly, on the time-scale of~$\sim 100 t_0$, consistent with the fact that the energy injection and dissipation rates nearly balance each other, see Fig.~\ref{fig:energy}. This suggests that the role of the imposed field and the associated magnetorotational instability in the steady-state cases I-III is merely to compensate for the very slight mismatch between the non-linear energy injection and dissipation rates. 

\edit{We also note that we were not able to observe} the MRI dynamo action in the $\mr{Pm}=1$ case \edit{despite a Reynolds number twice as large as in  \cite{fromang2007}.}

\section{Results}

In all the turbulent states the total energy spectra do not display good power-law scaling, see e.g., Fig.~\ref{fig:spectra}. This is consistent with previous studies~\cite[e.g.,][]{fromang2010,lesur2011}, where it was also found that while the total energy spectrum does not have good scaling, the kinetic spectrum exhibits the scaling somewhat close to~$k^{-3/2}$. 

\begin{figure}
  \resizebox{\columnwidth}{!} {\includegraphics{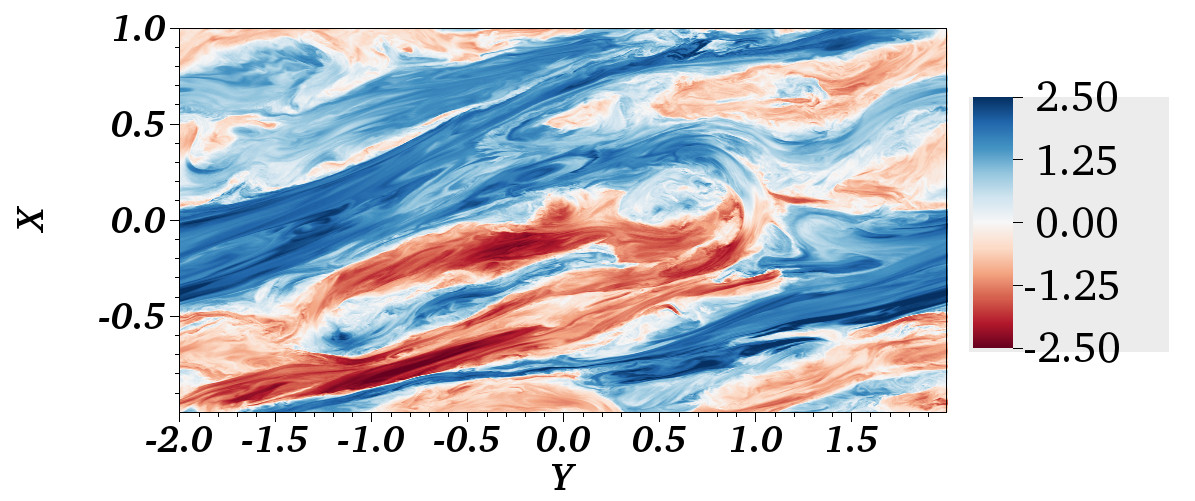}}
  \caption{Snapshot during case I of $b_y$, showing elongation of structures in $y$ direction and regions of strong, counter-aligned $b_y$ field.}
  \label{fig:snapshot}
\end{figure}

We find, however, that a more informative analysis can be performed if the field $b_y$ is separated from the total energy spectrum. As seen in Fig.~\ref{fig:spectra2}, the energy in $b_y$ is larger than in the rest of the fields; it is peaked at large scales and it rapidly declines with decreasing scale. Indeed, due to the $\Omega$-effect, fluctuations of magnetic field in a sheared flow become more aligned with the shear, enhancing the strength of the field in the shear direction, see Fig.~\ref{fig:snapshot}. The energy spectrum of the~$b_y$ fields scales closely to $k^{-2}$ which is possibly related to the domain structure seen in Fig.~\ref{fig:snapshot}, with sharp boundaries between the domains where the direction of the field reverses. The large-scale field $b_y$ may be responsible for the energy supply through the MRI instability, which is also concentrated at large scales, see Fig.~\ref{fig:injection}. 

The field $b_y$ plays the role of a background ``guiding" field for the remaining small-scale fluctuations, whose energy spectrum is close to~$k^{-3/2}$ in the interval \edit{$k \approx 4-20$}. Indeed, this scaling is consistent with the inertial range of large-scale, driven, steady-state, MHD turbulence\footnote{In the studies of driven MHD turbulence \cite[e.g.,][]{perez_etal2012}, one typically uses the Reynolds number based on the velocity fluctuations. In our case~I, this would give $\mr{Re}_{\mr {rms}} = v_{\mr {rms}} (L_z/2\pi)/\nu \approx 4000$.}~\cite[e.g.,][]{maron_g01,haugen_04,muller_g05,mininni_p07,chen_11,mason_cb06,mason_cb08,perez_b10_2,perez_etal2012,chandran_14}. This was found in both the non-zero and zero net-flux cases. The energy spectrum Fig.~\ref{fig:spectra2} is shown for case~I, and almost identical plots could be produced for cases~II and~IV -- case~III had much less energy and, therefore, a quite limited inertial range, see Table~\ref{tab:cases}. This suggests that the observed spectral behaviour is independent of a net flux or of the overall level of turbulence, and that, instead, it is an inherent property of the shearing, rotating flow.

\begin{table}
  \caption{Steady-state cases I, II, and III, and decaying case IV. \edit{The last two columns show best fits of the exponents of $E(k)-0.5\, b^2_y(k) \propto k^{-\zeta_\perp}$ and $0.5\, b^2_y(k) \propto k^{-\zeta_\parallel}$ over the range $k \in [4,20]$. For case~IV, the spectra are computed for $t/t_0 \in [20,50]$.}}   
  \label{tab:cases}
  \centering
  \begin{tabular}{c| c| c| c| c| c| c| c}
    \hline
    \hline
    Case & $B_0$ & $E$ & $\tilde{\alpha}$ & $\epsilon/E$ & \edit{$\zeta_\perp$} & \edit{$\zeta_\parallel$}\\
    \hline
    I   & 0.030 & 0.71  & 0.37 & 0.52 & \edit{$1.50 \pm 0.03$} & \edit{$1.95 \pm 0.02$} \\
    II  & 0.010 & 0.41  & 0.22 & 0.54 & \edit{$1.51 \pm 0.03$} & \edit{$1.96 \pm 0.03$} \\
    III & 0.005 & 0.072 & 0.039 & 0.54 & \edit{$1.13 \pm 0.08$} & \edit{$1.45 \pm 0.12$} \\
    IV  & 0.0 & \edit{-} & \edit{-} & 0.54 & \edit{$1.48 \pm 0.04$} & \edit{$1.95 \pm 0.04$} \\
    \hline
  \end{tabular}
\end{table}

Further insight in the MRI-driven turbulence can be gained from case~IV, where no magnetic flux is imposed, and, in  our case of $\mr{Pm}=1$, turbulence intensity declines. The decline is, however, very slow, on the order of~$\sim 100t_0$; so the system is observed to go through a sequence of quasi-steady states. In these states, as seen in Fig.~\ref{fig:energy}, the energy injection rate~${\tilde \alpha}$ nearly balances the rate of energy dissipation~$\epsilon$, and, therefore, the rate of energy cascade due to turbulence. The scaling of the turbulence spectrum does not practically change during this evolution, while both the turbulent energy~$E$ and the energy injection rate~${\tilde \alpha}$ slowly decay with time. An interesting property of such evolution is that the ratios $\epsilon/E$ and ${\tilde \alpha}/\epsilon$ remain nearly constant, as seen in Fig.~\ref{fig:energy} and in Table~\ref{tab:cases}. This indicates that the energy cascade time at the outer scale of turbulence is the same for all the observed quasi-steady states. 
\begin{figure}
\includegraphics{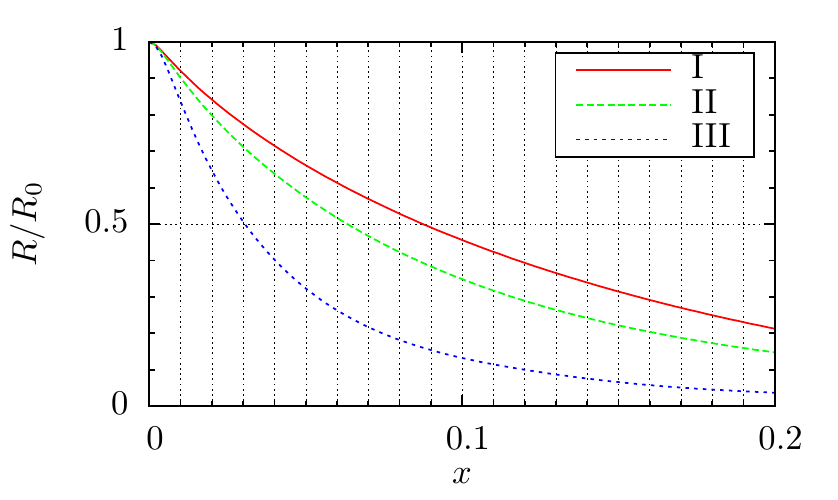}
\includegraphics{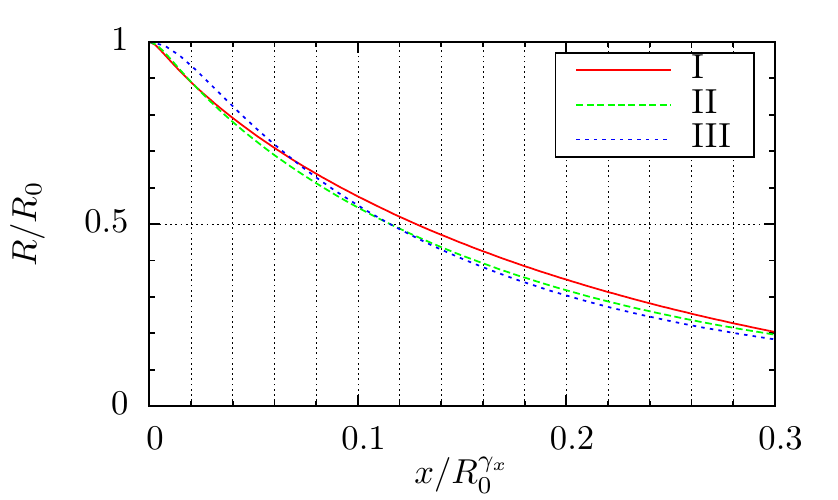}
\caption{Upper panel: correlation functions for steady-state cases~I, II~and~III, each scaled to its respective maximum. Lower panel: the $x$-axis is rescaled by $R_0^{\gamma_x}$, where $\gamma_x \approx 0.43$ minimizes the distance between the curves.}
\label{fig:corrs}
\end{figure}

A similar behaviour is observed in the steady-state, MRI-driven cases, which we now analyse in greater detail. We compare the three steady-state cases I-III, whose parameters and results are summarised in Table~\ref{tab:cases}. Similarly to the decaying case, we observe that while energy, dissipation and injection rates all change with the the imposed flux, the ratios $\epsilon/E$ and ${\tilde \alpha}/\epsilon$  remain constant. For a quantitative analysis of this phenomenon, in Fig.~\ref{fig:corrs} we plot the auto-correlation function of the fluctuations $R(x) = \langle\vec{v}(\vec{r} +  x\ex) \bm{\cdot} \vec{v}(\vec{r}) + \vec{b}(\vec{r} +  x\ex) \bm{\cdot} \vec{b}(\vec{r})\rangle$. The width of the auto-correlation function gives the typical scale (i.e., the outer scale) of the fluctuations, while its amplitude $R_0 = \langle v^2 \rangle + \langle b^2 \rangle$~\footnote{$\langle \cdot \rangle$ here and in the definition of $R(x)$ indicates a spatial average over each snapshot and an additional average over several snapshots.} gives their typical strength.

We observe from the lower panel of Fig.~\ref{fig:corrs} that the auto-correlation functions become remarkably similar if their spatial scales are renormalized  by~$R_0^{\gamma_x}$. The best fit is given by $\gamma_x \approx 0.43$, which is close to the value  $\gamma_x=0.5$ expected for the constant non-linear interaction rate of large-scale turbulent fluctuations. Indeed, a simple phenomenological consideration estimates this rate as~$\sim v_0/\lambda_0$, where $v_0=R_0^{1/2}$ is the intensity of fluctuations. A slight discrepancy between the two scalings may be related to the limited Reynolds numbers currently available to our analysis.  For larger Reynolds  numbers, we expect the form of the renormalized correlation function to be largely independent of Reynolds number.

We propose the following phenomenological explanation for this observation. Under the sole action of the orbital shear in equations (\ref{eq:mom}, \ref{eq:induct}) the energy is supplied to the system and transferred in the direction of large wavenumbers. Under the sole action of non-linear interaction the energy is removed from all the scales by a turbulent cascade, lowering the energy at the peak wavenumber. The rate of non-linear interaction increases with the wavenumber while the orbital-shear rate remains constant. Therefore, the orbital shear dominates at small $k$, while the non-linear interaction at large $k$. At small wavenumbers where the orbital shear dominates, the energy is shifted in the phase space toward large~$k$. This shift continues until the wavenumbers are reached where the rate of non-linear interaction competes with the orbital-shear rate and the energy is removed from large scales by a turbulent cascade. Therefore the scale $\lambda_0$ where the orbital shear is comparable to the rate of non-linear interaction, ${v_0}/{\lambda_0}\sim r\left({\mr d\Omega}/{\mr dr}\right)$, becomes the outer scale of the resulting turbulence. 

This condition constrains possible turbulent states in the shearing box. According to our consideration, the particular state may depend on the parameters of the system, such as the net magnetic flux, as we see in Table~\ref{tab:cases}, however, all such states satisfy the constraint formulated above.\footnote{\edit{It may seem that more detailed study of this phenomenology could be performed by varying the orbital shearing rate. We note, however, that varying the shearing rate {\em alone}, without changing other dimensionless parameters of the system, is a nontrivial task. Changing the shearing rate would imply changing the Rossby number (ratio of shear to rotation) which in turn would change the intensity of the resulting turbulence \citep[e.g.,][]{pessah2008}.}} 

The similarity and the universality of behaviour of small scales in MRI turbulence and in driven MHD turbulence lend support to the suggestion in~\cite[][]{fromang2007} that the higher Rm-number threshold for the dynamo action observed in the low Pm-number, magnetorotational case may be related to the similar effect observed in driven, isotropic turbulence. Analytic  consideration~\cite[e.g.,][]{boldyrev2004,boldyrev2010} and numerical simulations~\cite[e.g.,][]{iskakov2007} suggest that the turbulent magnetic dynamo action has a higher-threshold magnetic Reynolds number in low-Prandtl-number systems.  While in the case of magnetohydrodynamic turbulence the Rm-threshold value saturates as~Pm decreases \cite[e.g.,][]{kraichnan1967,vainshtein1986,boldyrev2004,boldyrev2010}, it remains to be seen whether a similar behaviour holds in the magnetorotational case. 

Our simulations \edit{show that the MRI dynamo is still nonexistent for $\mathrm{Rm}=45\,000$. This Reynolds number is about 8 times larger than the critical Reynolds number required for the dynamo action at $\mr{Pm}=4$ \citep{fromang2007}. This indicates that the MRI dynamo action at $\mr{Pm}\leq 1$, if possible at all, is much more difficult to obtain than the magnetic dynamo action  in isotropic, non-rotating turbulence, for which the threshold at $\mr{Pm}\ll 1$ is only about $3$~times higher compared to the threshold at~$\mr{Pm}>1$ \citep{iskakov2007}.}

\section{Conclusions}

The shearing-box model provides a simplified but highly nontrivial description of local turbulence in shearing, rotating flows. The mathematical properties of MRI driven turbulence of such a ``minimal'' model are not fully understood. Based on numerical simulations, we have proposed a phenomenological picture of the MRI-driven turbulence. We have shown that 1- the spectrum of MRI turbulence is independent of the mean field and may be understood in the framework of standard driven MHD turbulence, 2- the outer scale of MRI turbulence adjusts so that the turnover time is a constant fraction of the large scale shear, and \edit{3- the MRI dynamo action does not exist at $\mathrm{Pm=1}$ for $\mathrm{Rm}$ as high as 45\,000.} We believe that these invariant features will be the founding principles of a future predictive model for MRI turbulence.

\section*{Acknowledgements}
This work was supported by the US DoE grant DE-SC0003888 and the NSF Center for Magnetic Self-organization in Laboratory and Astrophysical Plasmas at U. Wisconsin-Madison. This research was supported in part by the National Science Foundation under Grant No. NSF PHY11-25915; GL and SB appreciate the hospitality and support of the Kavli Institute for Theoretical Physics, University of California, Santa Barbara, where part of this work was conducted. Simulations were performed at the Texas Advanced Computing Center (TACC) at the University of Texas at Austin under the NSF-Teragrid Project TG-PHY110016. SB is also supported by the Space Science Institute.

\bibliographystyle{mnras}
\bibliography{mri}

\begin{thebibliography}{}
\makeatletter
\relax
\def\mn@urlcharsother{\let\do\@makeother \do\$\do\&\do\#\do\^\do\_\do\%\do\~}
\def\mn@doi{\begingroup\mn@urlcharsother \@ifnextchar [ {\mn@doi@}
  {\mn@doi@[]}}
\def\mn@doi@[#1]#2{\def\@tempa{#1}\ifx\@tempa\@empty \href
  {http://dx.doi.org/#2} {doi:#2}\else \href {http://dx.doi.org/#2} {#1}\fi
  \endgroup}
\def\mn@eprint#1#2{\mn@eprint@#1:#2::\@nil}
\def\mn@eprint@arXiv#1{\href {http://arxiv.org/abs/#1} {{\tt arXiv:#1}}}
\def\mn@eprint@dblp#1{\href {http://dblp.uni-trier.de/rec/bibtex/#1.xml}
  {dblp:#1}}
\def\mn@eprint@#1:#2:#3:#4\@nil{\def\@tempa {#1}\def\@tempb {#2}\def\@tempc
  {#3}\ifx \@tempc \@empty \let \@tempc \@tempb \let \@tempb \@tempa \fi \ifx
  \@tempb \@empty \def\@tempb {arXiv}\fi \@ifundefined
  {mn@eprint@\@tempb}{\@tempb:\@tempc}{\expandafter \expandafter \csname
  mn@eprint@\@tempb\endcsname \expandafter{\@tempc}}}

\bibitem[\protect\citeauthoryear{{Balbus} \& {Hawley}}{{Balbus} \&
  {Hawley}}{1991}]{balbus1991}
{Balbus} S.~A.,  {Hawley} J.~F.,  1991, \mn@doi [\apj] {10.1086/170270}, \href
  {http://adsabs.harvard.edu/abs/1991ApJ...376..214B} {376, 214}

\bibitem[\protect\citeauthoryear{{Balbus} \& {Hawley}}{{Balbus} \&
  {Hawley}}{1998}]{balbus1998}
{Balbus} S.~A.,  {Hawley} J.~F.,  1998, \mn@doi [Reviews of Modern Physics]
  {10.1103/RevModPhys.70.1}, \href
  {http://adsabs.harvard.edu/abs/1998RvMP...70....1B} {70, 1}

\bibitem[\protect\citeauthoryear{{Balbus} \& {Henri}}{{Balbus} \&
  {Henri}}{2008}]{balbus2008}
{Balbus} S.~A.,  {Henri} P.,  2008, \mn@doi [\apj] {10.1086/524838}, \href
  {http://adsabs.harvard.edu/abs/2008ApJ...674..408B} {674, 408}

\bibitem[\protect\citeauthoryear{{Bodo}, {Mignone}, {Cattaneo}, {Rossi}  \&
  {Ferrari}}{{Bodo} et~al.}{2008}]{bodo2008}
{Bodo} G.,  {Mignone} A.,  {Cattaneo} F.,  {Rossi} P.,   {Ferrari} A.,  2008,
  \mn@doi [\aap] {10.1051/0004-6361:200809730}, \href
  {http://adsabs.harvard.edu/abs/2008A%26A...487....1B} {487, 1}

\bibitem[\protect\citeauthoryear{{Boldyrev} \& {Cattaneo}}{{Boldyrev} \&
  {Cattaneo}}{2004}]{boldyrev2004}
{Boldyrev} S.,  {Cattaneo} F.,  2004, \mn@doi [Physical Review Letters]
  {10.1103/PhysRevLett.92.144501}, \href
  {http://adsabs.harvard.edu/abs/2004PhRvL..92n4501B} {92, 144501}

\bibitem[\protect\citeauthoryear{{Chandran}, {Schekochihin}  \&
  {Mallet}}{{Chandran} et~al.}{2014}]{chandran_14}
{Chandran} B.~D.~G.,  {Schekochihin} A.~A.,   {Mallet} A.,  2014, preprint,
  \href {http://adsabs.harvard.edu/abs/2014arXiv1403.6354C} {} (\mn@eprint
  {arXiv} {1403.6354})

\bibitem[\protect\citeauthoryear{{Chandrasekhar}}{{Chandrasekhar}}{1960}]{chandra1960}
{Chandrasekhar} S.,  1960, \mn@doi [Proceedings of the National Academy of
  Science] {10.1073/pnas.46.2.253}, \href
  {http://adsabs.harvard.edu/abs/1960PNAS...46..253C} {46, 253}

\bibitem[\protect\citeauthoryear{{Chen}, {Mallet}, {Yousef}, {Schekochihin}  \&
  {Horbury}}{{Chen} et~al.}{2011}]{chen_11}
{Chen} C.~H.~K.,  {Mallet} A.,  {Yousef} T.~A.,  {Schekochihin} A.~A.,
  {Horbury} T.~S.,  2011, \mn@doi [MNRAS] {10.1111/j.1365-2966.2011.18933.x},
  \href {http://adsabs.harvard.edu/abs/2011MNRAS.415.3219C} {415, 3219}

\bibitem[\protect\citeauthoryear{{Fromang}}{{Fromang}}{2010}]{fromang2010}
{Fromang} S.,  2010, \mn@doi [\aap] {10.1051/0004-6361/201014284}, \href
  {http://adsabs.harvard.edu/abs/2010A%26A...514L...5F} {514, L5}

\bibitem[\protect\citeauthoryear{{Fromang}, {Papaloizou}, {Lesur}  \&
  {Heinemann}}{{Fromang} et~al.}{2007}]{fromang2007}
{Fromang} S.,  {Papaloizou} J.,  {Lesur} G.,   {Heinemann} T.,  2007, \mn@doi
  [\aap] {10.1051/0004-6361:20077943}, \href
  {http://adsabs.harvard.edu/abs/2007A%26A...476.1123F} {476, 1123}

\bibitem[\protect\citeauthoryear{{Gissinger}, {Ji}  \& {Goodman}}{{Gissinger}
  et~al.}{2011}]{gissinger2011}
{Gissinger} C.,  {Ji} H.,   {Goodman} J.,  2011, \mn@doi [\pre]
  {10.1103/PhysRevE.84.026308}, \href
  {http://adsabs.harvard.edu/abs/2011PhRvE..84b6308G} {84, 026308}

\bibitem[\protect\citeauthoryear{{Goldreich} \& {Lynden-Bell}}{{Goldreich} \&
  {Lynden-Bell}}{1965}]{goldreich1965}
{Goldreich} P.,  {Lynden-Bell} D.,  1965, \mnras, \href
  {http://adsabs.harvard.edu/abs/1965MNRAS.130..125G} {130, 125}

\bibitem[\protect\citeauthoryear{{Haugen}, {Brandenburg}  \& {Dobler}}{{Haugen}
  et~al.}{2004}]{haugen_04}
{Haugen} N. E.~L.,  {Brandenburg} A.,   {Dobler} W.,  2004, \mn@doi [Physical
  Review E] {10.1103/PhysRevE.70.016308}, \href
  {http://adsabs.harvard.edu/abs/2004PhRvE..70a6308H} {70, 016308}

\bibitem[\protect\citeauthoryear{{Hawley}, {Gammie}  \& {Balbus}}{{Hawley}
  et~al.}{1995}]{hawley1995}
{Hawley} J.~F.,  {Gammie} C.~F.,   {Balbus} S.~A.,  1995, \mn@doi [\apj]
  {10.1086/175311}, \href {http://adsabs.harvard.edu/abs/1995ApJ...440..742H}
  {440, 742}

\bibitem[\protect\citeauthoryear{{Herault}, {Rincon}, {Cossu}, {Lesur},
  {Ogilvie}  \& {Longaretti}}{{Herault} et~al.}{2011}]{herault2011}
{Herault} J.,  {Rincon} F.,  {Cossu} C.,  {Lesur} G.,  {Ogilvie} G.~I.,
  {Longaretti} P.-Y.,  2011, \mn@doi [\pre] {10.1103/PhysRevE.84.036321}, \href
  {http://cdsads.u-strasbg.fr/abs/2011PhRvE..84c6321H} {84, 036321}

\bibitem[\protect\citeauthoryear{{Iskakov}, {Schekochihin}, {Cowley},
  {McWilliams}  \& {Proctor}}{{Iskakov} et~al.}{2007}]{iskakov2007}
{Iskakov} A.~B.,  {Schekochihin} A.~A.,  {Cowley} S.~C.,  {McWilliams} J.~C.,
  {Proctor} M.~R.~E.,  2007, \mn@doi [Physical Review Letters]
  {10.1103/PhysRevLett.98.208501}, \href
  {http://adsabs.harvard.edu/abs/2007PhRvL..98t8501I} {98, 208501}

\bibitem[\protect\citeauthoryear{{Kagan} \& {Wheeler}}{{Kagan} \&
  {Wheeler}}{2014}]{kagan2014}
{Kagan} D.,  {Wheeler} J.~C.,  2014, \mn@doi [\apj]
  {10.1088/0004-637X/787/1/21}, \href
  {http://adsabs.harvard.edu/abs/2014ApJ...787...21K} {787, 21}

\bibitem[\protect\citeauthoryear{{Kraichnan} \& {Nagarajan}}{{Kraichnan} \&
  {Nagarajan}}{1967}]{kraichnan1967}
{Kraichnan} R.~H.,  {Nagarajan} S.,  1967, \mn@doi [Physics of Fluids]
  {10.1063/1.1762201}, \href
  {http://adsabs.harvard.edu/abs/1967PhFl...10..859K} {10, 859}

\bibitem[\protect\citeauthoryear{{Latter}, {Fromang}  \& {Faure}}{{Latter}
  et~al.}{2015}]{latter2015}
{Latter} H.~N.,  {Fromang} S.,   {Faure} J.,  2015, \mn@doi [\mnras]
  {10.1093/mnras/stv1890}, \href
  {http://adsabs.harvard.edu/abs/2015MNRAS.453.3257L} {453, 3257}

\bibitem[\protect\citeauthoryear{{Lesur} \& {Longaretti}}{{Lesur} \&
  {Longaretti}}{2007}]{lesur2007}
{Lesur} G.,  {Longaretti} P.-Y.,  2007, \mn@doi [\mnras]
  {10.1111/j.1365-2966.2007.11888.x}, \href
  {http://adsabs.harvard.edu/abs/2007MNRAS.378.1471L} {378, 1471}

\bibitem[\protect\citeauthoryear{{Lesur} \& {Longaretti}}{{Lesur} \&
  {Longaretti}}{2011}]{lesur2011}
{Lesur} G.,  {Longaretti} P.-Y.,  2011, \mn@doi [\aap]
  {10.1051/0004-6361/201015740}, \href
  {http://adsabs.harvard.edu/abs/2011A%26A...528A..17L} {528, A17}

\bibitem[\protect\citeauthoryear{{Lesur} \& {Ogilvie}}{{Lesur} \&
  {Ogilvie}}{2010}]{lesur2010a}
{Lesur} G.,  {Ogilvie} G.~I.,  2010, \mn@doi [\mnras]
  {10.1111/j.1745-3933.2010.00836.x}, \href
  {http://adsabs.harvard.edu/abs/2010MNRAS.404L..64L} {404, L64}

\bibitem[\protect\citeauthoryear{{Longaretti} \& {Lesur}}{{Longaretti} \&
  {Lesur}}{2010}]{longaretti2010}
{Longaretti} P.-Y.,  {Lesur} G.,  2010, \mn@doi [\aap]
  {10.1051/0004-6361/201014093}, \href
  {http://adsabs.harvard.edu/abs/2010A%26A...516A..51L} {516, A51}

\bibitem[\protect\citeauthoryear{{Malyshkin} \& {Boldyrev}}{{Malyshkin} \&
  {Boldyrev}}{2010}]{boldyrev2010}
{Malyshkin} L.~M.,  {Boldyrev} S.,  2010, \mn@doi [Physical Review Letters]
  {10.1103/PhysRevLett.105.215002}, \href
  {http://adsabs.harvard.edu/abs/2010PhRvL.105u5002M} {105, 215002}

\bibitem[\protect\citeauthoryear{{Maron} \& {Goldreich}}{{Maron} \&
  {Goldreich}}{2001}]{maron_g01}
{Maron} J.,  {Goldreich} P.,  2001, \mn@doi [Astrophys. J.] {10.1086/321413},
  \href {http://adsabs.harvard.edu/abs/2001ApJ...554.1175M} {554, 1175}

\bibitem[\protect\citeauthoryear{{Mason}, {Cattaneo}  \& {Boldyrev}}{{Mason}
  et~al.}{2006}]{mason_cb06}
{Mason} J.,  {Cattaneo} F.,   {Boldyrev} S.,  2006, \mn@doi [Physical Review
  Letters] {10.1103/PhysRevLett.97.255002}, \href
  {http://adsabs.harvard.edu/abs/2006PhRvL..97y5002M} {97, 255002}

\bibitem[\protect\citeauthoryear{{Mason}, {Cattaneo}  \& {Boldyrev}}{{Mason}
  et~al.}{2008}]{mason_cb08}
{Mason} J.,  {Cattaneo} F.,   {Boldyrev} S.,  2008, \mn@doi [Physical Review E]
  {10.1103/PhysRevE.77.036403}, \href
  {http://adsabs.harvard.edu/abs/2008PhRvE..77c6403M} {77, 036403}

\bibitem[\protect\citeauthoryear{{Mason}, {Perez}, {Boldyrev}  \&
  {Cattaneo}}{{Mason} et~al.}{2012}]{mason2012}
{Mason} J.,  {Perez} J.~C.,  {Boldyrev} S.,   {Cattaneo} F.,  2012, \mn@doi
  [Physics of Plasmas] {10.1063/1.3694123}, \href
  {http://adsabs.harvard.edu/abs/2012PhPl...19e5902M} {19, 055902}

\bibitem[\protect\citeauthoryear{{Meheut}, {Fromang}, {Lesur}, {Joos}  \&
  {Longaretti}}{{Meheut} et~al.}{2015}]{meheut2015}
{Meheut} H.,  {Fromang} S.,  {Lesur} G.,  {Joos} M.,   {Longaretti} P.-Y.,
  2015, \mn@doi [\aap] {10.1051/0004-6361/201525688}, \href
  {http://adsabs.harvard.edu/abs/2015A%26A...579A.117M} {579, A117}

\bibitem[\protect\citeauthoryear{{Mininni} \& {Pouquet}}{{Mininni} \&
  {Pouquet}}{2007}]{mininni_p07}
{Mininni} P.~D.,  {Pouquet} A.,  2007, \mn@doi [Physical Review Letters]
  {10.1103/PhysRevLett.99.254502}, \href
  {http://adsabs.harvard.edu/abs/2007PhRvL..99y4502M} {99, 254502}

\bibitem[\protect\citeauthoryear{{M{\"u}ller} \& {Grappin}}{{M{\"u}ller} \&
  {Grappin}}{2005}]{muller_g05}
{M{\"u}ller} W.~C.,  {Grappin} R.,  2005, \mn@doi [Physical Review Letters]
  {10.1103/PhysRevLett.95.114502}, \href
  {http://adsabs.harvard.edu/abs/2005PhRvL..95k4502M} {95, 114502}

\bibitem[\protect\citeauthoryear{{Perez} \& {Boldyrev}}{{Perez} \&
  {Boldyrev}}{2010}]{perez_b10_2}
{Perez} J.~C.,  {Boldyrev} S.,  2010, \mn@doi [Physics of Plasmas]
  {10.1063/1.3396370}, \href
  {http://adsabs.harvard.edu/abs/2010PhPl...17e5903P} {17, 055903}

\bibitem[\protect\citeauthoryear{{Perez}, {Mason}, {Boldyrev}  \&
  {Cattaneo}}{{Perez} et~al.}{2012}]{perez_etal2012}
{Perez} J.~C.,  {Mason} J.,  {Boldyrev} S.,   {Cattaneo} F.,  2012, \mn@doi
  [Physical Review X] {10.1103/PhysRevX.2.041005}, \href
  {http://adsabs.harvard.edu/abs/2012PhRvX...2d1005P} {2, 041005}

\bibitem[\protect\citeauthoryear{{Pessah} \& {Chan}}{{Pessah} \&
  {Chan}}{2008}]{pessah2008}
{Pessah} M.~E.,  {Chan} C.-k.,  2008, \mn@doi [\apj] {10.1086/589915}, \href
  {http://adsabs.harvard.edu/abs/2008ApJ...684..498P} {684, 498}

\bibitem[\protect\citeauthoryear{{Petitdemange}, {Dormy}  \&
  {Balbus}}{{Petitdemange} et~al.}{2008}]{petitdemange2008}
{Petitdemange} L.,  {Dormy} E.,   {Balbus} S.~A.,  2008, \mn@doi [\grl]
  {10.1029/2008GL034395}, \href
  {http://cdsads.u-strasbg.fr/abs/2008GeoRL..3515305P} {35, 15305}

\bibitem[\protect\citeauthoryear{{Rincon}, {Ogilvie}  \& {Proctor}}{{Rincon}
  et~al.}{2007}]{rincon2007}
{Rincon} F.,  {Ogilvie} G.~I.,   {Proctor} M.~R.~E.,  2007, \mn@doi [Physical
  Review Letters] {10.1103/PhysRevLett.98.254502}, \href
  {http://cdsads.u-strasbg.fr/abs/2007PhRvL..98y4502R} {98, 254502}

\bibitem[\protect\citeauthoryear{{Riols}, {Rincon}, {Cossu}, {Lesur},
  {Longaretti}, {Ogilvie}  \& {Herault}}{{Riols} et~al.}{2013}]{riols_etal2013}
{Riols} A.,  {Rincon} F.,  {Cossu} C.,  {Lesur} G.,  {Longaretti} P.-Y.,
  {Ogilvie} G.~I.,   {Herault} J.,  2013, \mn@doi [Journal of Fluid Mechanics]
  {10.1017/jfm.2013.317}, \href
  {http://adsabs.harvard.edu/abs/2013JFM...731....1R} {731, 1}

\bibitem[\protect\citeauthoryear{{Riols}, {Rincon}, {Cossu}, {Lesur}, {Ogilvie}
   \& {Longaretti}}{{Riols} et~al.}{2015}]{riols_etal2015}
{Riols} A.,  {Rincon} F.,  {Cossu} C.,  {Lesur} G.,  {Ogilvie} G.~I.,
  {Longaretti} P.-Y.,  2015, \mn@doi [\aap] {10.1051/0004-6361/201424324},
  \href {http://adsabs.harvard.edu/abs/2015A%26A...575A..14R} {575, A14}

\bibitem[\protect\citeauthoryear{{Seilmayer} et~al.,}{{Seilmayer}
  et~al.}{2014}]{seilmayer2014}
{Seilmayer} M.,  et~al., 2014, \mn@doi [Physical Review Letters]
  {10.1103/PhysRevLett.113.024505}, \href
  {http://adsabs.harvard.edu/abs/2014PhRvL.113b4505S} {113, 024505}

\bibitem[\protect\citeauthoryear{{Shakura} \& {Sunyaev}}{{Shakura} \&
  {Sunyaev}}{1973}]{shakura1973}
{Shakura} N.~I.,  {Sunyaev} R.~A.,  1973, \aap, \href
  {http://adsabs.harvard.edu/abs/1973A%26A....24..337S} {24, 337}

\bibitem[\protect\citeauthoryear{{Sisan}, {Mujica}, {Tillotson}, {Huang},
  {Dorland}, {Hassam}, {Antonsen}  \& {Lathrop}}{{Sisan}
  et~al.}{2004}]{sisan2004}
{Sisan} D.~R.,  {Mujica} N.,  {Tillotson} W.~A.,  {Huang} Y.-M.,  {Dorland} W.,
   {Hassam} A.~B.,  {Antonsen} T.~M.,   {Lathrop} D.~P.,  2004, \mn@doi
  [Physical Review Letters] {10.1103/PhysRevLett.93.114502}, \href
  {http://adsabs.harvard.edu/abs/2004PhRvL..93k4502S} {93, 114502}

\bibitem[\protect\citeauthoryear{{Umurhan} \& {Regev}}{{Umurhan} \&
  {Regev}}{2004}]{umurhan2004}
{Umurhan} O.~M.,  {Regev} O.,  2004, \mn@doi [\aap]
  {10.1051/0004-6361:20040573}, \href
  {http://adsabs.harvard.edu/abs/2004A%26A...427..855U} {427, 855}

\bibitem[\protect\citeauthoryear{{Vainshtein} \& {Kichatinov}}{{Vainshtein} \&
  {Kichatinov}}{1986}]{vainshtein1986}
{Vainshtein} S.~I.,  {Kichatinov} L.~L.,  1986, \mn@doi [Journal of Fluid
  Mechanics] {10.1017/S0022112086000290}, \href
  {http://adsabs.harvard.edu/abs/1986JFM...168...73V} {168, 73}

\makeatother
\end{thebibliography}

\end{document}